From: elisat@cmirl1.ge.infn.it
Message-Id: <9704220816.AA03594@cmirl1.ge.infn.it>
To: biferale@roma2.infn.it
Subject: ultima versione: vuoi controllare? 
Date: Tue, 22 Apr 97 10:16:11 +0200
X-Mts: smtp

domanda: sai se posso mandarlo cosi', cioe' in lpr style (2 colonne,ecc)
 oppure 
deve essere a una colonna, double spaced, ecc.....???
da cio' che leggo su internete mi pare che sia giusta la seconda,
ma tu mi avevi detto la prima..o sbaglio?

 \documentstyle[prl,aps,twocolumn,epsf]{revtex}

 \newcommand{\be}{\begin{equation}}
 \newcommand{\ee}{\end{equation}}
 \newcommand{\bdis}{\begin{displaymath}}
 \newcommand{\edis}{\end{displaymath}}
 \newcommand{\eg}{\varepsilon}

\begin{document}
 \bibliographystyle{prsty}

    \title{Ultrametric structure of multiscale energy correlations in
   turbulent models}
 \author{R.Benzi$^{1}$, L. Biferale$^{2}$
   and E. Trovatore$^{3}$}
  \address{$^{1}$  AIPA, Via Po 14, 00100 Roma, Italy,\\
  $^{2}$ Dipartimento di Fisica and INFM, Universit\`{a} 
  di Tor Vergata, Via della Ricerca Scientifica 1, I-00133 Roma, Italy,\\
  $^{3}$  CMIRL--Dipartimento di Fisica, Universit\`{a} 
 di Genova, Via Dodecaneso 33, I-16146, Genova, Italy.}
  \maketitle

  
\begin{abstract}
\\
Title:Ultrametric structure of multiscale energy correlations in
      turbulent models.
Authors: R. Benzi, L. Biferale and E. Trovatore
\\
Ultrametric structure of the energy cascade process in a 
dynamical model of turbulence is studied.
The tree model we use can be viewed as an approximated one--dimensional 
truncation of the wavelets resolved Navier--Stokes dynamics.
Varying the tree connectiveness, the appearance of a scaling transition 
in the two--points moments of energy dissipation is detected, 
in agreement with experimental turbulent data.
\\
\end{abstract}

\vspace{0.5cm}

 Spatio-temporal intermittency is the most intriguing aspect of a fully
developed three-dimensional turbulent flow.  
 
 Experiments \cite{MS}
show that the energy dissipation defines 
 a multifractal measure on the fluid volume. 
The multifractal measure is characterized
by the  scaling  properties of the coarse-grained
 energy dissipation on a box at scale $r$, $\eg_r$, namely:
 $<(\eg_r(x))^p> \sim r^{\tau(p)}$, 
where $<\cdot>$ means averaging over all boxes
 of size $r$ and centered in $x$ 
in which the volume can be partitioned.
The measured $\tau(p)$ exponents show a clear intermittent behaviour, 
i.e., a non-linear dependency on the 
order of the moment $p$. 

The simplest way to  explain phenomenologically  the presence
of intermittent deviations  consists in 
 describing the energy transfer mechanism in terms of fragmentation
stochastic processes. In these models 
(see \cite{logpoisson} for a recent proposal),
one  introduces a set of  eddies leaving on a dyadic  
structure and connected through a random multiplicative process.

Let us remark that all stochastic fragmentation models so far proposed
 lack any direct linking with the original
Navier-Stokes (NS) equations.
Dynamical deterministic models on  hierarchical structures 
are therefore invoked for 
improving our understanding of  the energy transfer mechanism.

In this paper, we study
 dynamical models which fill the gap between  purely
stochastic fragmentation models and the original  NS
dynamics.  
In particular, we consider a dynamical model
on  one-space and one-time dimensions \cite{BBTT}. 
  One can look at  this model 
 as  an approximation of the original NS
equations in a wavelets basis \cite{wave}.

In order to specify the model one has to select the set of
interactions connecting eddies at different scales and
at different spatial locations. 
By changing the interactions set, one changes the
scale--organization
of energy structures: in order to study it, new tools
are required \cite{OM93}, which 
characterize intermittency more completely than the multifractal spectrum
$\tau(p)$ alone.
An obvious generalization of the single--point statistics
is to inquire about the scaling of two--point moments.
In practice, introducing the mixed moments:

\be
<(\eg_r(x))^q  (\eg_l(x+s))^p>
\label{eq:twopoints}
\ee

one can study correlations between different scales by changing
$r$ and $l$ and/or correlations between different points in the fluid
volume by changing $s$.

Different  interactions
among nodes of the hierarchical dynamical structure can lead to
very different prediction for the scaling 
behaviour of (\ref{eq:twopoints}). 
An ultrametric organization of the tree
can be detected by looking for a phase transition in 
the set of scaling exponents characterizing 
correlations (\ref{eq:twopoints}).
In \cite{OM93} the analysis of the two-points
observables (\ref{eq:twopoints}) performed on experimental turbulent data
gave a first support for an ultrametric organization
of the main triadic interactions in Navier-Stokes eqs. 

In the following, we are going to analyse the same kind of observables
measured on a direct numerical integration of a dyadic-tree model
for turbulent energy cascade. In particular, 
by changing the set of interacting triads we want to 
disentangle the basic symmetries behind the transition
observed in
 real turbulent data.

Let us turn to a brief review of the model (for a comprehensive 
description see \cite{BBTT}).
The tree model can be viewed as an extension of shell models,
which can be seen as a severe truncation of the 
NS equations (see \cite{KA} for a general 
introduction).
The most popular shell model is  the Gledzer-Ohkitani-Yamada 
(GOY) model (\cite{G}--\cite{BKLMW}).
Recently, a new class of shell models based upon the helical
 decomposition of NS equations \cite{W}
has been  suggested \cite{BK} and studied \cite{BBKT,BBT}. 
In these models, one or few 
 complex variables $u_n$ represent an entire  
{\it shell} of wave numbers $k$ such that $k_n < k < k_{n+1}$,
with $k_n = 2^n $. 
Shell models can be thought of as  field 
problems in zero spatial dimension (d=0).
In order to include also some real space dynamics we need to transform 
 the {\it chain}-model 
 into a {\it tree}-model with $d=1$.
This is achieved by letting grow the number of degrees of freedom
with the shell index $n$ as $2^n$.
The tree model  can be regarded as 
describing the evolution of the coefficients of an orthonormal wavelets 
expansion of a one-dimensional projection of the 
velocity field.
In the tree model, we use
the notation $u^{\pm}_{n ,j}$ to indicate a complex 
variable having positive or negative defined helicity and living
on scale $k_n$ and spatial position labeled by the index 
$j$.
For a given shell $n$, the index $j$ can vary from $1$ to $2^{n-1}$.

We report here the structure of the tree model
 dynamical equations (for more details, see \cite{BBTT}):    

\begin{eqnarray}
\dot{u}^+_{n,j}&=&i k_n \sum_{n_1,n_2,j_1,j_2} 
[ a_{n_1,n_2,j_1,j_2} u^{s_1}_{n_1,j_1} u^{s_2}_{n_2,j_2}]^* \nonumber \\
&&-\nu k^2_n u^+_{n,j} +\delta_{n,n_0} F^+.
\label{eq:shells}
\end{eqnarray}

Here, $n=1,...,N$, where $N$ is the total
number of shells, $\nu$ is the viscosity, $F^+$ the external forcing acting
on a the large-scale shell $n_0=1$ and $a_{n_1,n_2,j_1,j_2}$ are parameters, 
which  are determined by imposing
conservation of energy and  helicity in the inviscid and unforced limit.
The $s_1$ and $s_2$ indices are the helicity signs ($\pm$) of interacting
modes. The same equations hold, with all helicities reversed, 
for $u^-_{n,j}$.

In restricting the possible choices of the nonlinear terms,
we can
 phenomenologically  require
a certain degree of locality for interactions among variables at different 
scales and at different spatial positions.
Regarding scale numbers $n_1$ and $n_2$, 
 in our equations (\ref{eq:shells})
 each variable $u^+_{n,j}$ is 
allowed to interact only with nearest and next--to--nearest levels:
indeed, $n_1$ and $n_2$ can vary only from $(n-2)$ to $(n+2)$.
Regarding space numbers $j_1$ and $j_2$,
we define two different models having different topological
structure of the dynamical interactions. The first model, hereafter
called model A, is a model which has an {\it ultrametric}
structure. That is,  each eddy is allowed to 
interact only with bigger
eddies which spatially contain it and
with smaller eddies  
spatially contained  in it. This model is the natural dynamical
representative of all stochastic fragmentation models which
phenomenologically reproduce single-point intermittent exponents.
The second model, hereafter called model B, 
has an  enlarged interactions set containing
also horizontal couplings, which allow eddies covering 
different spatial regions to
interact each other.
In figures 1 and 2, we pictorially show the set of interactions
defining models A and B.

The single--point statistical properties of the tree
model have been studied in 
\cite{BBTT}: in both cases A and B,
the system  turned out to have
an intermittent energy transfer 
 qualitatively  similar to what one can find in the original NS eqs.
  The  tree-like structure 
imposed on the velocity fluctuations does not necessary implies
 that the energy dissipation can be described in terms of
fragmentation processes. In order to test the scale-organization
of the energy structures, ultrametric-sensitive observables should
be studied. 

In order to compare our system with previous findings \cite{OM93},
 we shall detect possible ultrametric structure in the energy 
cascade  of the tree model looking at
two--point statistical quantities.
All parameters settings and numerical methods are as in \cite{BBTT}. 
In particular, we consider a total number of levels 
$N=16$: the  total number of sites forming the tree is then $N_T = 2^N-
1=65535$,
each one  described in terms of two complex variables.
Numerical simulations needed state--of--the--art multi--processor
computers.

The  fields we focus on  are the coarse-grained energy
 dissipation densities, here denoted as
$\epsilon_n(j)$, 
obtained as averages over spatial regions $\Lambda_j(n)$
of length $2^{-n}$.
We consider the mixed moments  (\ref{eq:twopoints}) with 
$r=l$ and $p=-q$,
which in our notation become:

\be
<\eg_n^q(j) \eg_n^{-q}(j+s)> \equiv C_n(q,s)
\label{eq:corre}
\ee

The behaviour of this quantity for intermittent ultrametric 
 measures  resulting from
random multiplicative processes has been already analyzed in the framework 
of the two--point multifractal formalism \cite{OM93,CD87}.
In this case, the average can be properly decomposed 
and a general result
can be obtained for its dependence on the spatial distance $s$
between the two points:

\be
C_n(q,s) \sim s^{min[-\tau(-q)-\tau(q),1]} \equiv s^{\Phi(q)}.
\label{eq:corr}
\ee 

This expression implies that for some moment $q$, 
  a sharp transition occurs in the derivative of the scaling exponent
$\Phi(q)$.
This scaling transition is the analog of
 a phase transition in the thermodynamic
interpretation of multifractals \cite{FJP86}.
The behaviour of (\ref{eq:corre}) is dominated by
 pairs of points at which the
dissipation is very large at one point and very low at the other.
This constitutes the subset of points that are likely to be independent 
from each other and lie on the boundary of their bigger predecessors.
Indeed, it must be recalled that in a ultrametric structure
nearby (in space)  eddies could lie on the boundaries
of much bigger ones, then having an effective  large ultrametric distance.

The two spatial scales of interest are 
the coarse--graining scale $l_n=2^{-n}$ and 
the offset scale $l_s=l_n s$: they  should be such that 
$\eta \ll l_n \ll l_s \ll \Lambda_T$, where $\eta$ and 
$\Lambda_T$ are the Kolmogorov and integral scale, respectively.
For this reason, we fixed $n=11$, in order to consider the largest inertial
scale in our tree structure, and we let $s$ to assume the 
exponentially spaced values $s=2^m$, with $m=1,2,...,n-2$.
In order to test the presence of a scaling transition, the mixed moments
(\ref{eq:corre}) 
have been computed for
$0.5 < q < 4$.

Figure 3 shows the mixed moments (\ref{eq:corr}) as a function 
of $s$ in log--log coordinates, for increasing values of $q$ and for
the two versions A and B of the tree model.
The $\Phi(q)$ exponents have been calculated by linear fit in the 
inertial range region: they are reported
in figure 4,  where are
compared with the curve $(-\tau(q)-\tau(-q))$, obtained using 
the single--point moments exponents and corresponding to
the predicted form of $\Phi(q)$ if the scaling transition were absent.
In the case A, the data support a sharp
transition in the derivative of $\Phi(q)$ 
at $q \sim 1.5$, with a much slower variation of $\Phi(q)$
for $q > 1.5$.
This transition is absent in case B.

We thus conclude that version A gives support for 
 a scaling transition in the mixed moments of coarse--grained dissipation.
This result agrees with the experimental behaviour
  found in \cite{OM93} (see fig.17 of this reference) using
data measured in a turbulent wake.
The physical picture implied by this scaling transition
is that of uncorrelated small eddies that come close 
together even sharing no common history during the energy cascade.

Let us summarize our results. 

A dynamical model in one spatial dimension
originating from a wavelets-like decomposition of a one-dimensional
 cut of a turbulent velocity field has been studied.
We found that a scaling transition appears as soon
as the tree has a pure ultrametric dynamical structure.

The fact that decreasing the number of 
triad interactions one can reproduce  the real data scaling transition
observed in \cite{OM93}  may seem in contrast with 
the observation that in the original Navier-Stokes equations 
all possible interactions are switched on. This
contradiction is only apparent: divergenceless character
of the original NS field, added to complex phase-coherence effects,
can   very easily introduce different dynamical weights in the possible
 triad interactions  leading to a situation
where only a few of them govern the global dynamical evolution.
For example, Grossmann and coworkers showed \cite{grossmann1,grossmann2},
 by performing 
suitable truncation of NS equations,
that intermittency depends on the typical  degree
of locality  in Fourier space of the survived triad interactions;
very similar results have also been found in shell models 
at varying the inter-shell ratio $\lambda$ \cite{BBT}.

Recently, some theoretical studies and experimental analysis
have been done on multi--point multi-scale velocity correlation
functions in turbulent flows \cite{proc1}.
Our dynamical investigation suggests
that,  in the presence of a strong
ultrametric structure, correlations among velocity fluctuations
at different scales should depend on their ultrametric distance 
rather than on the separation length only, as predicted in \cite{proc1}. 

These and similar studies performed on
such kind of models can improve our understanding of basic mechanisms
underlying  turbulent cascade.
For instance, it is important to recognize
 those interactions
which are more effective in the energy transfer mechanism
 when constructing eddy-viscosity models
and in simulating  small scale
statistics by some
 closure approach.


  
  \vspace{1cm}

 {\large{\bf Figures Captions}}
 
 \vspace{1cm}
 
 \begin{itemize}
 
 \item[-] {Figure 1: Pictorial representation of  interactions 
     in model A: in this case 
	the nonlinear terms in (\protect\ref{eq:shells})
	result from the sum of parts (a), (b) and (c)  
	in the figure.}
 
 \item[-] {Figure 2: Pictorial representation of  interactions 
     in model B: in this case
	the nonlinear terms in (\protect\ref{eq:shells})
	result from the sum of parts (a), (b), (c), (d), (e)
	and (f) 
	in the figure.}
 
 \item[-] {Figure 3: Log--log plot of mixed moments of order
 $q=0.5,1,1.5,...,4$ (from bottom
to top) against the distance $s$, for model A (left)and B 
(right).}

 \item[-] {Figure 4: The exponents $\Phi(q)$ (black circles)
as a function of $q$ for
model A (left) and B (right).
For comparison, the values of the function $(-\tau(q)-\tau(-q))$
are also reported (white circles).}

\end{itemize}
  

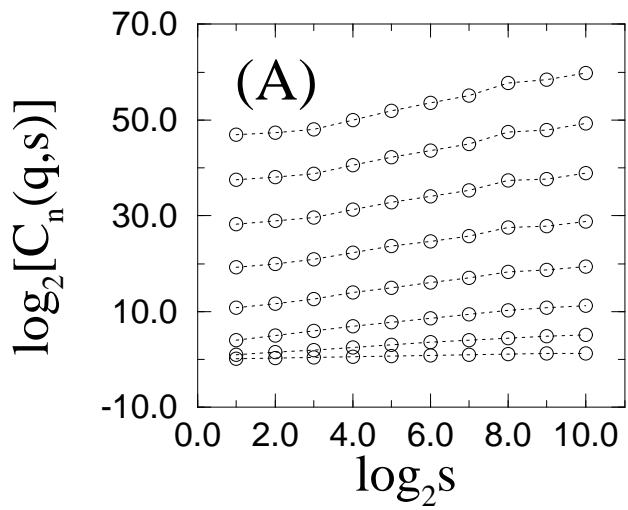
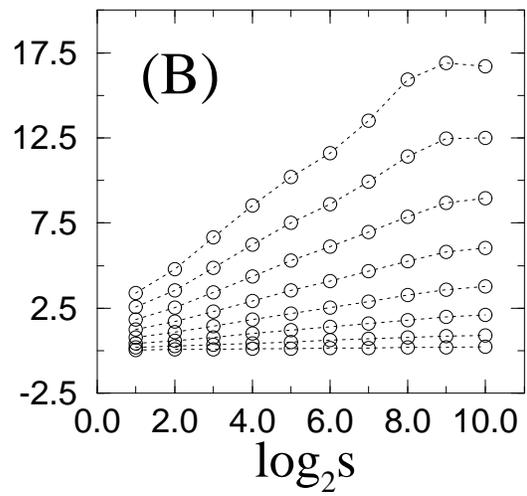
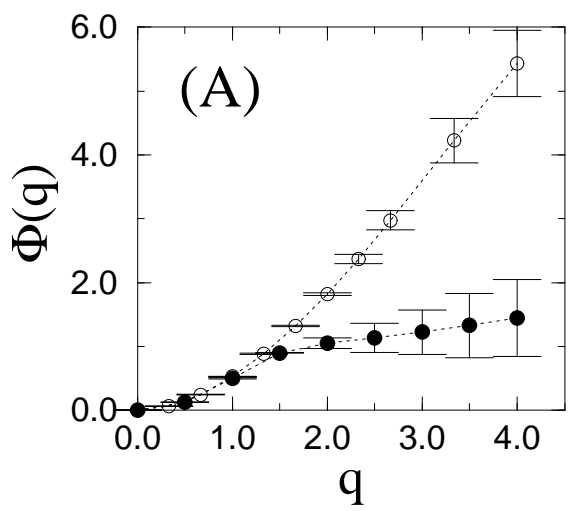
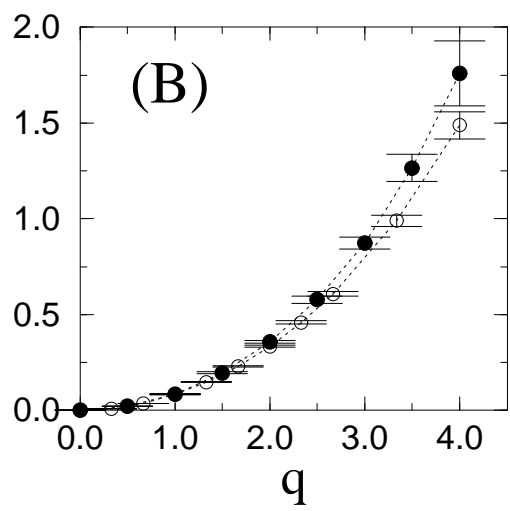
\end{document}